\documentclass{article}
\usepackage{amssymb}
\usepackage{amsmath}
\usepackage{graphicx}

\input{tcilatex}
\begin{document}

\begin{center}
{\LARGE Sagnac delay in the Kerr-dS space-time: Implications for Mach's
principle}

\bigskip

\bigskip

R.Kh. Karimov$^{1,a}$, R.N. Izmailov$^{1,b}$, G.M. Garipova$^{2,c}$ and K.K.
Nandi$^{1,2,d}$,

\bigskip

$^{1}$Zel'dovich International Center for Astrophysics, Bashkir State
Pedagogical University, 3A, October Revolution Street, Ufa 450000, RB, Russia

$^{2}$Department of Physics \& Astronomy, Bashkir State University, 47A,
Lenin Street, Sterlitamak 453103, RB, Russia

$\bigskip $

$^{a}$E-mail: karimov\_ramis\_92@mail.ru

$^{b}$E-mail: izmailov.ramil@gmail.com

$^{c}$E-mail: goldberg144@gmail.com

$^{d}$E-mail: kamalnandi1952@yahoo.co.in

---------------------------------------------------------------

\bigskip

\textbf{Abstract}
\end{center}

Relativistic twin paradox can have important implications for Mach's
principle. It has been recently argued that the behavior of the time
asynchrony (different aging of twins) between two flying clocks along closed
loops can be attributed to the existence of an absolute spacetime, which
makes Mach's principle unfeasible. In this paper, we shall revisit, and
support, this argument from a different viewpoint using the Sagnac delay.
This is possible since the above time asynchrony is known to be exactly the
same as the Sagnac delay between two circumnavigating light rays re-uniting
at the orbiting source/receiver. We shall calculate the effect of mass $M$
and cosmological constant $\Lambda $ on the delay in the general case of
Kerr-de Sitter spacetime. It follows that, in the independent limits $%
M\rightarrow 0$, spin $a\rightarrow 0$ and $\Lambda \rightarrow 0$, while
the Kerr-dS metric reduces to Minkowski metric, the clocks need \textit{not}
tick in consonance since there will still appear a non-zero observable
Sagnac delay. While we do not measure spacetime itself, we do measure the
Sagnac effect, which signifies an absolute substantive Minkowski spacetime
instead of a void. We shall demonstrate a completely different limiting
behavior of Sagnac delay, heretofore unknown, between the case of
non-geodesic and geodesic source/observer motion.

\begin{center}
---------------------------------------------------------------

\textbf{1. Introduction}
\end{center}

\baselineskip=4ex%
Mach's principle has been interpreted in so numerously equivalent ways that
depending on the interpretation, even opposite conclusions could be reached.
Mach opposed Newton's concept of abstract absolute space determined by
"fixed stars" but advocated that inertial forces should be caused by
accelerations with respect to all other "bodies" in the universe (relational
program). It is one version of Mach's principle. Another variant is that
static sources should cause static spacetimes. However, this version has not
been adequately incorporated in Einstein's general relativity (GR) theory:
The G\"{o}del universe rotates even though the source stress tensor is
static, which showed that Mach's principle is not preserved in GR. A
genuinely Machian theory of gravity was developed by Brans and Dicke (BD) in
1961 but observationally there is little difference with Einstein's GR at
least in the weak field solar system tests since the BD coupling parameter\
is quite large.

Lichtenegger and Iorio [1] recently argued in the context of the twin
paradox and Mach's principle that\ the different aging of the twins need not
conform to Machian ideas but the ultimate cause of the behavior of the
clocks must be attributed to the independent status of spacetime. Machian
idea of the relativity of motion requires that spacetime should loose its
metric properties in a universe devoid of all mass-energy. They argued that,
in the limit of zero mass-energy, the spacetime does not dissolve into
nothingness but becomes Minkowskian producing observable effects, which
makes Mach's relational program unfeasible. Our analysis will strengthen
this argument from a different viewpoint.

The different aging of twins is no paradox but a result of an absolute
synchronization discontinuity that occurs between two clocks flying in
closed loop around a spinning mass, say, Earth. If two identical clocks on
the equatorial plane depart from a point with equal speeds relative to
Earth, one eastward and the other westward, they will have equal energies
but not equal time rates leading to a synchronization discontinuity between
the two flying clocks when they reunite (meaning twins aging differently).
Schlegel [2] has shown that this clock synchronization discontinuity is
exactly the same as the Sagnac delay, and argued that only when this delay
is taken into account in the empirical formula, the Hafele-Keating around
the Earth experiment [3] confirms the time dilation effect of special
relativity. However, even though the Sagnac delay is caused by spinning
Earth (frame dragging), it is neither a mass nor velocity dependent effect
to zeroth order, and thus has absolute character.

The Sagnac delay breifly is as follows: Consider a circular turntable of
radius $R$ having a light source/receiver (meaning the source \textit{and}
the receiver at the same point) fixed on the turntable\footnote{%
The fixed position on the turntable of the source/observer cannot be its
geodesic motion since it will be acted on by the artificial forces, unlike
for example Earth's orbital motion, which being geodesic or in free fall to
Sun, such forces are balanced by gravitational pull (Weak Equivalence
Principle).}. A beam of light split into two at the source/receiver are made
to follow the same closed path along the rim in opposite directions before
they are re-united at the source/receiver. If the turntable is not rotating,
the beams will arrive at the same time at the source/receiver and an
interference fringe will appear. When the turntable rotates with angular
velocity $\omega _{0}$, the arrival times at the source/receiver will be
different for co-rotating and counter-rotating beams: longer in the former
case and shorter in the latter. This difference in arrival times is called
the Sagnac delay (named after the discoverer) [4]. The\ total arrival time
lag between the two light beams, as measured at the source/receiver for its
east and westward motion, can be obtained from special relativity, which
gives, to first order in $\omega _{0}$, 
\begin{equation}
\delta \tau _{S}=\frac{4\omega _{0}S}{c^{2}},
\end{equation}%
where $S$ ($=\pi R^{2}$) is the area enclosed by the equatorial circular
orbit, orthogonal to the rotation axis,\ of the closed path followed by the
waves contouring the turntable, $c$ is the speed of light in vacuum. The
effect has been previously investigated in different solutions of Einstein
general relativity (see, e.g., [5-9]). It is possible to move ahead from
special relativity and consider general relativistic corrections to the
Sagnac delay (1), when the "turntable" is a massive spinning compact object
like the Earth. Note that there is no mass term in Eq.(1), so we expect to
recover it from the zeroth order (flat space) part of the total general
relativistic delay.

The purpose of the present paper is to first briefly show how Sagnac delay
can result even in the Minkowski space written in rotating coordinates. Then
we shall assume the Kerr-dS spacetime to represent the gravitational field
of a compact spinning mass at large distances. The reason for chosing this
spacetime is that it is more general than just the Kerr spacetime and that
asymptotically it is de Sitter, not Minkowski, representing repulsive dark
energy supported by current cosmological observations. We shall consider
both non-geodesic and geodesic circular motions and expose the differences
reflected in the Sagnac delay. Then we shall recover the delay (1) in the
zeroth order post-Newtonian approximation in the non-geodesic case. The
geodesic motion will also be computed and discussed.

The paper is organized as follows: In Sec.2, we briefly reproduce how Sagnac
delay can result in the flat Minkowski space written in a coordinate system
rotating uniformly. In Sec.3, we present the Kerr-dS metric and Secs.4 and 5
are devoted to the Sagnac delay for non-geodesic circular motion and post
Newtonian approximation respectively. Sec.6 deals with the delay in the case
of geodesic circular motion. Sec.7 concludes the paper. We shall take units
such that $G=1$, $c=1$, unless they are specifically restored.

\begin{center}
\textbf{2. Sagnac delay in the Minkowski spacetime}
\end{center}

To derive Eq.(1) in the flat space, we consider the Minkowski metric in
cylindrical coordinates%
\begin{equation}
ds^{2}=c^{2}dt^{\prime 2}-\left( dr^{\prime 2}+r^{\prime 2}d\phi ^{\prime
2}+dz^{\prime 2}\right) 
\end{equation}%
and change over to a uniformly spinning coordinate system ($t,r,\phi ,z$)
given by%
\begin{equation}
r^{\prime }=r,\phi ^{\prime }=\phi +\omega _{0}t,z^{\prime }=z,ct^{\prime
}=ct=x^{0},
\end{equation}%
then the metric (2) changes into%
\begin{equation}
ds^{2}=(c^{2}-\omega _{0}^{2}r^{2})dt^{2}-2\omega _{0}r^{2}d\phi
dt-dr^{2}-r^{2}d\phi ^{2}-dz^{2},
\end{equation}%
where $\omega _{0}$ is the angular velocity of the spinning system with
respect to the static intertial Minkowski space. This is a well known
uniformly spinning flat metric (\textit{metric on the turntable}), where
test particles are acted on by artificial forees. This metric has been used
for exemplifying concepts like gravitational red shift, frame dragging,
Lense-Thirring effect etc.[10-12]. Note that the metric (4) is valid only
out to radial distances $r<c/\omega _{0}$, beyond which $g_{00}$ becomes
negative, which is not permitted. 

We recall the most general form of the metric%
\begin{equation}
ds^{2}=g_{00}\left( dx^{0}\right) ^{2}+2g_{0i}dx^{0}dx^{i}+g_{ij}dx^{i}dx^{j}%
\text{, \ }i.j=1,2,3
\end{equation}%
and note that, in a rotating coordinate system, the time gap $dx^{0}$
between two simultaneous events separated by an infinitesimal distance $%
dx^{i}$ is (see, e.g., [13] for details)%
\begin{equation}
dx^{0}=-\frac{g_{0i}dx^{i}}{g_{00}}\equiv g_{i}dx^{i}.
\end{equation}%
Therefore, for a closed loop trajectory in the metric (5), it follows that%
\begin{equation}
\Delta t=-\frac{1}{c}\doint \frac{g_{0i}dx^{i}}{g_{00}}=\frac{1}{c^{2}}%
\doint \frac{\omega _{0}r^{2}}{1-\left( \omega _{0}r/c\right) ^{2}}d\phi .
\end{equation}%
Assuming $\omega _{0}r/c<<1$, which also ensures that the proper time 
\begin{equation}
\Delta \tau =\Delta t\sqrt{1-\left( \omega _{0}r/c\right) ^{2}}\simeq \Delta
t,
\end{equation}%
we get from the above%
\begin{equation}
\Delta \tau \simeq \frac{\omega _{0}}{c^{2}}\doint r^{2}d\phi =\pm \frac{%
2\omega _{0}S}{c^{2}},
\end{equation}%
where $S$ ($\equiv \pi R^{2}$) is the area enclosed by the circular
trajectory of radius $R$ and the $+$ refers to motion along the spin and the 
$-$ sign opposite to the spin respectively. Therefore, two light rays
proceeding simultaneously along $\pm $ directions from a fixed point
(source/receiver) at a distance $R$ from the center of the spinning
spacetime will reunite at that point with a total proper time gap 
\begin{equation}
\delta \tau =\frac{4\omega _{0}S}{c^{2}}.
\end{equation}

This is just the Sagnac delay (1), which also appears in the empirical
Hafele and Keating formula [3] (neglecting Earth's gravitational potential
effect):%
\begin{equation}
\Delta t^{\prime }\simeq \left( 1-\frac{v^{2}}{2c^{2}}-\frac{vR\omega _{0}}{%
c^{2}}\right) \Delta t,
\end{equation}%
where $\Delta t$ is the time interval on the clock at rest on Earth and $%
\Delta t^{\prime }$ is the time of the clock flying along the Earth's
equator with speed $v$ along $\pm $ directions, $\omega _{0}$ is the Earth's
axial rotational velocity and $R$ is the radius of the Earth. Schlegel [2]
has shown that, with $\Delta t=2\pi R/v$, the third term yields a time gap%
\begin{equation}
\pm \frac{vR\omega _{0}}{c^{2}}\Delta t=\pm \frac{2\pi \omega _{0}R^{2}}{%
c^{2}}=\pm \frac{2\omega _{0}S}{c^{2}},
\end{equation}%
which is just the same as Eq.(9)\footnote{%
Equation (1) represents a clock synchronization discontinuity that has been
empirically confirmed in the Hafele-Keating experiment, so it is also called
Hafele-Keating discontinuity [2].}. When this term is factored out from
(11), one ends up with the confirmation of the special relativistic time
dilatation formula. Note that (12) is independent of $v$ and Earth's mass $M$%
, thus has an absolute character. We shall derive it as the zeroth order
term from the delay in general relativistic Kerr-dS metric.

\begin{center}
\textbf{3. Kerr - dS metric }
\end{center}

The Kerr-dS metric (also called Carter's metric [14], see also [15,16]) is a
spinning solution of Einstein's field equations%
\begin{equation}
R_{\mu \nu }=\Lambda g_{\mu \nu },
\end{equation}%
where $\Lambda >0$ is the cosmological constant representing repulsive dark
energy density. The full metric is given by 
\begin{equation}
d\tau ^{2}=\frac{\Delta _{r}}{\rho ^{2}\Xi ^{2}}\left[ dt-a\text{sin}^{2}{%
\theta }d\phi \right] ^{2}-\frac{\Delta _{\theta }\text{sin}^{2}{\theta }}{%
\rho ^{2}\Xi ^{2}}\left[ (r^{2}+a^{2})d\phi -adt\right] ^{2}-\frac{\rho ^{2}%
}{\Delta _{r}}dr^{2}-\frac{\rho ^{2}}{\Delta _{\theta }}d\theta ^{2},
\end{equation}%
where for convenience we are redefining $\Lambda \equiv 6\gamma $ so that 
\begin{equation}
\Delta _{r}=(r^{2}+a^{2})\left( 1-\gamma r^{2}\right) -2Mr,
\end{equation}%
\begin{equation}
\rho ^{2}=r^{2}+a^{2}\text{cos}^{2}{\theta },
\end{equation}%
\begin{equation}
\Delta _{\theta }=1+\gamma a^{2}\text{cos}^{2}{\theta },
\end{equation}%
\begin{equation}
\Xi =1+\gamma a^{2},\text{ \ }
\end{equation}%
where $M$ is the (asymptotic) mass of the source, $a$ is the ratio between
the angular momentum $J$ and the mass $M,$ 
\begin{equation}
a=\frac{J}{M}.
\end{equation}%
When $\gamma =0$, one recovers the usual Kerr solution in Boyer-Lindquist
coordinates. We shall compute the Sagnac delay for two types of equatorial
orbits in the ensuing sections and consider only weak field effects allowing
corresponding approximations to be made.

\begin{center}
\textbf{4. Sagnac delay for non-geodesic circular equatorial orbit}
\end{center}

We shall follow the method developed by Tartaglia [8]. Consider that the
source/receiver, sending two oppositely directed light beams, is orbiting
around a spinning compact object described by metric (14), along a
circumference on the equatorial plane $\theta =\pi /2$. Suitably placed
mirrors send back to their origin both beams after a circular trip about the
central mass. Assume further that source/receiver is orbiting at a radius $%
r=R=$ const. sharing a uniform rotational velocity $\omega _{0}$ of the
central spinning source. Then the metric (14) reduces to 
\begin{equation*}
d\tau ^{2}=\frac{R^{2}-2MR+a^{2}-\gamma R^{2}(R^{2}+a^{2})}{%
R^{2}(1+a^{2}\gamma )^{2}}(dt-ad\phi )^{2}
\end{equation*}%
\begin{equation}
-\frac{1}{R^{2}(1+a^{2}\gamma )^{2}}[(R^{2}+a^{2})d\phi -adt]^{2}.
\end{equation}%
The rotation angle $\phi _{0}$ of the source/receiver is 
\begin{equation}
\phi _{0}=\omega _{0}t.
\end{equation}%
Then 
\begin{equation}
d\tau ^{2}=\frac{R^{2}[1-(R^{2}+a^{2})\{\omega _{0}^{2}+(a\omega
_{0}-1)^{2}\gamma \}]-2MR(a\omega _{0}-1)^{2}}{R^{2}(1+a^{2}\gamma )^{2}}%
dt^{2}.
\end{equation}%
For light moving along the same circular path, $d\tau =0$. Assuming $\Omega $
to be the angular velocity of light rays along the path, we have 
\begin{equation}
R^{2}[1-(R^{2}+a^{2})\{\Omega ^{2}+(a\Omega -1)^{2}\gamma \}]-2MR(a\Omega
-1)^{2}=0,\text{ }a^{2}\gamma \neq -1,
\end{equation}%
Solving Eq.(23), one finds two roots that represent the angular velocity $%
\Omega _{\pm }$ of light for the co- and counter rotating light motion given
by 
\begin{equation*}
\Omega _{\pm }=\frac{2aM/R+a(R^{2}+a^{2})\gamma }{R^{2}+2(M/R)a^{2}+a^{2}%
\left\{ 1+(R^{2}+a^{2})\gamma \right\} }
\end{equation*}%
\begin{equation}
\pm \frac{\sqrt{R^{2}-2MR+a^{2}-R^{2}(R^{2}+a^{2})\gamma }}{%
R^{2}+2(M/R)a^{2}+a^{2}\left\{ 1+(R^{2}+a^{2})\gamma \right\} }.
\end{equation}%
The rotation angles $\phi _{\pm }$ for light are then 
\begin{equation}
\phi _{\pm }=\Omega _{\pm }t.
\end{equation}%
Eliminating $t$ between Eqs.(21) and (25), we obtain 
\begin{equation}
\phi _{\pm }=\frac{\Omega _{\pm }}{\omega _{0}}\phi _{0}.
\end{equation}%
The first intersection of the world lines of the two light rays with the one
of the orbiting source/receiver after the emission at time $t=0$ is, when
the angles are 
\begin{equation}
\phi _{+}=\phi _{0}+2\pi ,
\end{equation}%
\begin{equation}
\phi _{-}=\phi _{0}-2\pi ,
\end{equation}%
which give

\begin{equation}
\frac{\Omega _{\pm }}{\omega _{0}}\phi _{0}=\phi _{0}\pm 2\pi .
\end{equation}%
Solving for $\phi _{0}$, 
\begin{equation}
\phi _{0\pm }=\mp \frac{2\pi \omega _{0}}{\Omega _{\pm }-\omega _{0}},
\end{equation}%
we have, putting the expressions from (24), 
\begin{equation*}
\phi _{0\pm }=\mp 2\pi \omega _{0}/\left[ \frac{2aM/R+a(R^{2}+a^{2})\gamma }{%
R^{2}+2(M/R)a^{2}+a^{2}\{1+(R^{2}+a^{2})\gamma \}}\right. 
\end{equation*}%
\begin{equation}
\left. \pm \frac{\sqrt{R^{2}-2MR+a^{2}-R^{2}(R^{2}+a^{2})\gamma }}{%
R^{2}+2(M/R)a^{2}+a^{2}\left\{ 1+(R^{2}+a^{2})\gamma \right\} }-\omega _{0}%
\right] .
\end{equation}%
The proper time of the orbiting source/receiver, deduced from Eq.(22) using
Eq.(21), is 
\begin{equation}
d\tau =\frac{\sqrt{R^{2}[1-(R^{2}+a^{2})\{\omega _{0}^{2}+(a\omega
_{0}-1)^{2}\gamma \}]-2MR(a\omega _{0}-1)^{2}}}{R(1+a^{2}\gamma )}\frac{%
d\phi _{0}}{\omega _{0}}.
\end{equation}%
Finally, integrating between $\phi _{0-}$ and $\phi _{0+}$ , we obtain the
exact Sagnac delay 
\begin{equation}
\delta \tau =\frac{\sqrt{R^{2}[1-(R^{2}+a^{2})\{\omega _{0}^{2}+(a\omega
_{0}-1)^{2}\gamma \}]-2MR(a\omega _{0}-1)^{2}}}{R(1+a^{2}\gamma )}\frac{\phi
_{0+}-\phi _{0-}}{\omega _{0}}.
\end{equation}%
Using the integration limits from Eq.(31), we explicitly write the exact
value as 
\begin{equation*}
\delta \tau =\frac{4\pi }{R}\left[ \left\{
R^{3}+2Ma^{2}+a^{2}R+a^{2}R(R^{2}+a^{2})\gamma \right\} \omega
_{0}-2Ma\right. 
\end{equation*}%
\begin{equation*}
\left. -aR(R^{2}+a^{2})\gamma \right] /\left[ (1+a^{2}\gamma
)\{1-2M/R+4a(M/R)\omega _{0}\right. 
\end{equation*}%
\begin{equation}
\left. -(R^{2}+2Ma^{2}/R+a^{2})\omega _{0}^{2}-(a\omega
_{0}-1)^{2}(R^{2}+a^{2})\gamma \}^{1/2}\right] .
\end{equation}%
The delay (34) is often interpreted as the gravitational analogue of the
Bohm-Aharonov effect [17] although the light beams are not truly moving in
the gravitation free space. The best situation that possibly comes closer to
the Bohm-Aharonov effect could be developed with light beams moving along a
flat space torus (see for details, Semon [18]). Nevertheless, as shown by
Ruggiero [19], expression (34) completely agrees with the one of the
gravito-electromagnetic Bohm-Aharonov interpretation [20]. For the viewpoint
of Bohm-Aharonov quantum interference in general relativity, see [21,22].

On the other hand, we can imagine a source/receiver keeping a fixed position
in a coordinate system defined by distant fixed stars ($\omega _{0}=0$). For
him, a Sagnac delay will also occur under the condition that $a\neq 0$,
given by 
\begin{equation}
\delta \tau _{0}=-\frac{8\pi a\{M+\gamma R(a^{2}+R^{2})/2\}}{R(1+a^{2}\gamma
)\sqrt{1-2M/R-(a^{2}+R^{2})\gamma }}.
\end{equation}%
A Post-Newtonian first order approximation for a static observer sending a
pair of light beams in opposite directions along a closed triangular
circuit, instead of a circle, was worked out by Cohen and Mashhoon [9] and
they found the same result as above in that approximation. So what is
important is not the shape but the closedness of the orbit.

\begin{center}
\textbf{5. Post-Newtonian approximation}
\end{center}

Eq.(34) is the exact result for the Sagnac delay for the equatorial motion.
In most cases many terms in this equation are very small allowing series
approximations, which we do below. Let us first assume that $\beta =\omega
_{0}R\ll 1$, and develop Eq.(34) in powers of $\beta $ retaining terms only
up to the second order. The result is 
\begin{equation*}
\delta \tau \simeq -\frac{8\pi a\left\{ M+\gamma R(a^{2}+R^{2})/2\right\} }{%
R(1+a^{2}\gamma )\sqrt{1-2M/R-(a^{2}+R^{2})\gamma }}
\end{equation*}%
\begin{equation*}
+\frac{4\pi \left\{ R^{2}-2MR+a^{2}-R^{2}(a^{2}+R^{2})\gamma \right\} }{%
R(1+a^{2}\gamma )\left\{ 1-2M/R-(a^{2}+R^{2})\gamma \right\} ^{3/2}}\beta 
\end{equation*}%
\begin{equation}
-\frac{12\pi a\left[ \{M+\gamma
R(a^{2}+R^{2})/2\}\{1-2MR+a^{2}/R^{2}-(a^{2}+R^{2})\gamma \}\right] }{%
R(1+a^{2}\gamma )\left\{ 1-2M/R-(a^{2}+R^{2})\gamma \right\} ^{5/2}}\beta
^{2},
\end{equation}%
which displays that the first term is just $\delta \tau _{0}$ of Eq.(35), as
expected. Now we perform a successive post-Newtonian approximation in $%
\varepsilon =M/R\ll 1$ and in $a/R$ $\ll 1$, and using the expression $%
\delta \tau _{S}=4\omega _{0}S=4\omega _{0}\pi R^{2}=4\pi \beta R$, we
obtain the final result 
\begin{equation*}
\delta \tau \simeq \delta \tau _{S}\left\{ 1+\frac{\gamma R^{2}}{2}-\gamma
a^{2}\left( 1+\frac{\gamma R^{2}}{2}\right) \right\} 
\end{equation*}%
\begin{equation*}
+4\pi RM\omega _{0}\left\{ 1+\frac{3\gamma R^{2}}{2}-\gamma a^{2}\left( 1+%
\frac{3\gamma R^{2}}{2}\right) \right\} 
\end{equation*}%
\begin{equation}
-\frac{8\pi aM}{R}\left\{ 1+\gamma R^{2}-\gamma a^{2}\left( 1+\gamma
R^{2}\right) \right\} .
\end{equation}%
The flat space Sagnac effect $\delta \tau _{S}$ is \textit{not} completely
recovered even when the correction terms containing $M$ and $a$ are
negligible, due to the appearance of an extra non-local term $\frac{\gamma
R^{2}}{2}$ due to $\gamma $, viz.,%
\begin{equation}
\delta \tau \simeq \delta \tau _{S}\left\{ 1+\frac{\gamma R^{2}}{2}\right\} .
\end{equation}%
When, in addition, $\gamma =\Lambda /6\rightarrow 0$, we recover the zeroth
order Sagnac delay that coincides with the delay in Minkowski spacetime
derived in Sec.2. 
\begin{equation}
\delta \tau \simeq \delta \tau _{S}=\frac{4\omega _{0}S}{c^{2}}.
\end{equation}

\begin{center}
\textbf{6. Sagnac delay for geodesic circular equatorial orbit}
\end{center}

The previous equatorial orbit was not geodesic or in free fall since the
source/receiver was moving with rotational velocity $\omega _{0}$ not
required to satisfy Kepler's third law. Here we are considering a circular
geodesic orbit of the source/receiver (maybe a free fall satellite) at some
arbitrary radius on the equator ($\theta =\pi /2$) and sending light signals
circumnavigating the Earth. The rotational velocity $\omega _{\pm }$ of the
satellite is now determined by the circular geodesic itself.

Defining the velocity four-vector $\overset{.}{x}^{\nu }=\frac{dx^{\nu }}{%
d\tau }$, the Lagrangian can be written as%
\begin{equation}
L=\frac{1}{2}g_{\mu \nu }\overset{.}{x}^{\mu }\overset{.}{x}^{\nu }
\end{equation}%
and the Euler-Lagrange $r-$equation is%
\begin{equation}
\frac{d}{d\tau }\left( \frac{\partial L}{\partial \overset{.}{r}}\right) =%
\frac{\partial L}{\partial r}.
\end{equation}%
Since in metric (14), $g_{r\mu }=0$ for $r\neq \mu $, we have%
\begin{equation}
\frac{d}{d\tau }\left( g_{rr}\overset{.}{r}\right) =\frac{1}{2}g_{\mu \nu ,r}%
\overset{.}{x}^{\mu }\overset{.}{x}^{\nu }.
\end{equation}%
Circular orbits are defined by the conditions%
\begin{equation}
\overset{.}{r}=\overset{..}{r}=0,
\end{equation}%
so that the Eq.(42) yields%
\begin{equation}
g_{tt,r}\overset{.}{t}^{2}+2g_{t\phi ,r}\overset{.}{t}\overset{.}{\phi }%
+g_{\phi \phi ,r}\overset{.}{\phi }^{2}=0.
\end{equation}%
Defining $\omega =\overset{.}{\phi }/\overset{.}{t}$, this equation yields
the quadratic equation \ 
\begin{equation}
g_{\phi \phi ,r}\omega ^{2}+2g_{t\phi ,r}\omega +g_{tt,r}=0.
\end{equation}%
\ 

From the metric (14), putting $dr=0$ at $r=R=$ const. and $d\theta =0$ at $%
\theta =\pi /2$, we find%
\begin{equation*}
d\tau ^{2}=g_{tt}dt^{2}+2g_{t\phi }dtd\phi +g_{\phi \phi }d\phi ^{2}\text{,}
\end{equation*}%
where%
\begin{eqnarray}
g_{tt} &=&1-\frac{2M}{R}-\gamma \left( a^{2}+R^{2}\right) ,\text{ }g_{t\phi
}=\frac{2aM}{R}+\gamma a\left( a^{2}+R^{2}\right) ,  \notag \\
g_{\phi \phi } &=&-\frac{2a^{2}M}{R}-\left( a^{2}+R^{2}\right) \left(
1+a^{2}\gamma \right) .
\end{eqnarray}%
The source/receiver rotational velocities $\omega _{\pm }$ then follow from
the two roots of Eq.(45), using Eqs.(46), 
\begin{equation}
\omega _{\pm }=\frac{\left( \frac{aM}{R^{2}}-aR\gamma \right) \pm \sqrt{%
\frac{M}{R}-\gamma R^{2}}}{\frac{a^{2}M}{R^{2}}-R-a^{2}R\gamma }.
\end{equation}%
Putting it into $\delta \tau _{S\pm }=4\pi R^{2}\omega _{\pm }$,, we obtain
the exact delay. One could treat this result as representing the effect of
cosmological constant $\Lambda $ on the Sagnac delay. When $a=0$, $\gamma =0$%
, we have $\omega _{\pm }=\mp \sqrt{\frac{M}{R^{3}}}$, which is just
Kepler's third law. We now expand Eq.(47) up to first order in $\left(
a/R\right) $ and obtain 
\begin{equation}
\omega _{\pm }=\left( \gamma R-\frac{M}{R^{2}}\right) \left( \frac{a}{R}%
\right) \pm \frac{1}{R}\sqrt{\frac{M}{R}-\gamma R^{2}}.
\end{equation}%
Noting that $\omega _{\pm }$ $=$ const. (since $r=R=$ const. for circular
orbits), we can insert it into the first order delay\ to obtain $\delta \tau
_{S\pm }=4\pi R^{2}\omega _{\pm }$, so that 
\begin{equation}
\delta \tau _{S\pm }=\pm 4\pi \left[ \sqrt{MR-\gamma R^{4}}\mp \left( \frac{a%
}{R}\right) \left( M-\gamma R^{3}\right) \right] 
\end{equation}%
The Kerr terms follow at $\gamma =0$, when we recover the formula in
Lichtenegger and Iorio [1]: 
\begin{equation}
\delta \tau _{\pm }=\pm 4\pi \sqrt{MR}\mp 4\pi a\left( \frac{M}{R}\right) .
\end{equation}

\begin{center}
\textbf{7. Conclusions}
\end{center}

We wish to clarify a crucial point: The idea was to start with the Kerr-dS
metric due to a spinning mass $M\neq 0$, and then expand the exact Sagnac
delay obtaining a zeroth order term at $M\rightarrow 0$, which has been
shown to be\ just the usual Sagnac delay (1) that cannot be made to vanish
by any coordinate transformation. However, physically, the massive turntable
cannot be made massless, and even if this happens, the source/observer will
fly away along tangential directions destroying the circular orbit.
Therefore, the premise of our argument is \textit{not} to let the mass
vanish but sift out observable effects, if any, that pertains only to the
asymptotic Minkowski spacetime. Sagnac delay is precisely the effect the
Hafele-Keating experiment has tested, which also contains the special
relativistic time dilation that depends on clock speed $v$ although the
Sagnac correction is independent of $M$ and $v$. Independence from mass does
not mean vanishing of mass, the limit $M\rightarrow 0$ is only a formal way
to sift out the measurable delay in the flat space, and nothing more.
Therefore, we think that synchronization discontinuity or different aging of
twins or Sagnac delay are the same thing and the flat space effect shows
that Mach's relational program can indeed be unfeasible.

In detail, we calculated the exact Sagnac delay in the general relativitic
Kerr-dS spacetime assuming that in the asymptotic limit it adequately
describes the spacetime of any compact spinning object, not necessarily a
black hole. This allowed us to make a post-Newtonian expansion of the exact
delay for both non-geodesic and geodesic circular source/observer motions.
The zeroth order post-Newtonian approximation (39) has been confirmed by the
Hafele-Keating experiment [3] in the round the Earth clock experiment. This
can be seen when the Earth values $\omega _{0}=$ $\omega _{\oplus
}=7.30\times 10^{-5}$ rad/s and $R_{\oplus }=6,378,137$ m, when put in (39)
yield the famous result $\Delta \tau _{S}\simeq 2\times 207.4$ ns. This
measurable zeroth order effect reinforces the conclusion of Lichtenegger and
Iorio [1] that the ultimate cause of the behavior of the clocks (different
aging of twins) must be attributed to the absolute status of spacetime (here
Minkowski) that is in contravention to the relational program of Mach.

We have found that the two types of closed loop motions produce completely
different types of behavior of the Sagnac delay. For the non-geodesic
circular source/observer motion in the Kerr-dS metric, the exact Sagnac
delay in Eq.(34) shows the influence of $\Lambda $ on it. (The motion is
called non-geodesic since the angular velocity $\omega _{0}$ does not obey
Kepler's third law).  This is in line with other similar works in the
literature, where the influence of $\Lambda $ on gravitomagnetic [23], light
deflection [24-27] and perihelion advance [28], time delay and other effects
[29] were derived. From Eq.(37), we showed that the limits $M\rightarrow 0,$ 
$a\rightarrow 0$ and $\Lambda \rightarrow 0$ yield just the delay in
Minkowski spacetime shown in Sec.2. This means that, on reuniting at the
source/observer after circumnavigation, the clocks will show synchronization
discontinuity or different aging of twins, that is, clocks not ticking at
the same rate. This observable synchronization discontinuity (like
international dateline) is an absolute property of the spinning Minkowski
metric alone\footnote{%
We wish to point out that a similar type of limiting process has been
considered by Chakraborty and Majumdar [30] (see also [31]). They calculated
an exact expression for the Lense-Thirring effect for Kerr-Taub-NUT
spacetime and showed that in the case of the zero angular momentum ($%
a\rightarrow 0$) Taub-NUT spacetime, the frame-dragging effect does not
vanish, when considered for spinning test gyroscopes. They regarded this
residual Lense-Thirring frequency as an effect of \textit{Copernican frame}
(which we think to be just the absolute frame of Newton determined by fixed
stars).}, which is contrary to Mach's principle.

For the geodesic circular motion, we calculated in Sec.6 the angular
velocities $\omega _{\pm }$ of the source/observer that yielded Kepler's law 
$\omega _{\pm }=\sqrt{\frac{M}{R^{3}}}$ in the limit $a\rightarrow 0$, $%
\Lambda \rightarrow 0$ (Schwarzschild limit). Unlike Eq.(34), Eq.(49) now
contains the gravitating mass $M$ in the leading order and the limits $%
M\rightarrow 0,$ $a\rightarrow 0$ and $\Lambda \rightarrow 0$ simply yield $%
\delta \tau _{S\pm }=0$ in the Minkowski spacetime. This happens because $%
\omega _{\pm }=0$, or no circular motion. In this case, at least the Sagnac
effect does not appear as an observable in the Minkowski limit, and Mach's
principle could be valid. The above results throw light on the different
behaviour of Sagnac delay depending on the types of source/observer motion
that does not seem to have been noticed heretofore..

\textbf{Acknowledgment}

Part of the work was supported by the Russian Foundation for Basic Research
(RFBR) under Grant No.16-32-00323.

\textbf{References}

[1] H.I.M. Lichtenegger and L. Iorio, Eur. Phys. J. Plus \textbf{126}, 129
(2011).

[2] R. Schlegel, Nature (London), \textbf{242}, 180 (1973).

[3] J.C. Hafele and R.E. Keating, Science \textbf{177}, 166 (1972), \textit{%
ibid. }\textbf{177}, 168 (1972).

[4] G. Sagnac, C. R. Acad. Sci. Paris, \textbf{157}, 708 (1913).

[5] A. Bhadra, T.B. Nayak and K.K. Nandi, Phys. Lett. A\textbf{\ 295}, 1
(2002).

[6] K.K. Nandi, P.M. Alsing, J.C. Evans and T.B. Nayak, Phys. Rev. D\textbf{%
\ 63}, 084027 (2001).

[7] A. Ashtekar and A. Magnon, J. Math. Phys. \textbf{16}, 343 (1975).

[8] A. Tartaglia, Phys. Rev. D\textbf{\ 58}, 064009 (1998).

[9] J. M. Cohen and B. Mashhoon, Phys. Lett. A \textbf{181}, 353 (1993).

[10] A. Einstein, \textit{The Meaning of Relativity}, Princeton U.P.,
Princeton, NJ, (1955), pp. 55--63.

[11] B. F. Schutz, \textit{A First Course in General Relativity, }Cambridge
U.P., New York (1985), p. 298.

[12] T. A. Weber, Am. J. Phys. \textbf{65}, 486 (1997).

[13] L. D. Landau and E. M. Lifshitz, \textit{The Classical Theory of Fields}%
, Pergamon, New York, 4th ed., (1975), pp. 234--237, 248--258.

[14] B. Carter, in \textit{Les Astres Occlus} eds. C. DeWitt, \& B. DeWitt
(Gordon \& Breach, New York (1973).

[15] J.A.R. Cembranos, A. de la Cruz-Dombriz, P. Jimeno Romero, Int. J.
Geom. Meth. Mod. Phys. \textbf{11}, 1450001 (2014).

[16] D. P\'{e}rez, G. E. Romero, and S.E. Perez Bergliaffa, Astron.
Astrophys. \textbf{551}, A4 (2013).

[17] Y. Aharonov and D. Bohm, Phys. Rev. \textbf{115}, 485 (1959).

[18] M. D. Semon, Found. Phys. \textbf{12}, 49 (1982).

[19] M. L. Ruggiero, Nuovo Cim. B \textbf{119}, 893 (2004).

[20] J. J. Sakurai, Phys. Rev. D \textbf{21}, 2993 (1980).

[21] K.K. Nandi and Y.-Z. Zhang, Phys. Rev. D \textbf{66}, 063005 (2002).
\qquad

[22] P.M. Alsing, J.C. Evans and K.K. Nandi, Gen. Rel. Grav. \textbf{33},
1459 (2001).

[23] L. Iorio and M.L. Ruggiero, JCAP 0903 (2009) 024.

[24] W. Rindler and M. Ishak, Phys. Rev. D \textbf{76}, 043006 (2007).

[25] A. Bhattacharya, G.M. Garipova, E. Laserra, A. Bhadra and K.K. Nandi,
JCAP 02 (2011) 028.

[26] A. Bhattacharya, A. Panchenko, M. Scalia, C. Cattani and K.K. Nandi,
JCAP 09 (2010) 004.

[27] C. Cattani, M. Scalia, E. Laserra, I. Bochicchio and K.K. Nandi, Phys.
Rev. D \textbf{87}, 047503 (2013).

[28] M. Sereno and P. Jetzer, Phys. Rev. D \textbf{73}, 063004 (2006).

[29] V. Kagramanova, J. Kunz and C. L\"{a}mmerzahl, Phys. Lett. B \textbf{634%
}, 465 (2006).

[30] C. Chakraborty and P. Majumdar, Class. Quant. Grav. \textbf{31}, 075006
(2014).

[31] V. Kagramanova, J. Kunz, E. Hackmann and C. L\"{a}mmerzahl, Phys. Rev.
D \textbf{81}, 124044 (2010).

\begin{center}
---------------------------------------------------------------------
\end{center}

\bigskip

\bigskip

\bigskip

\bigskip

\end{document}